\documentclass[10pt]{article}

\usepackage[top=1in, bottom=1in, left=0.9in, right=0.9in]{geometry}
\usepackage{setspace}
\usepackage{makeidx}
\usepackage{graphicx}
\usepackage{textcomp}
\usepackage{float}
\usepackage{xcolor}
\usepackage{booktabs}
\usepackage{multirow}
\usepackage{array}
\usepackage{url}
\usepackage{tcolorbox}
\tcbset{
    colback=white,
    colframe=black,
    boxsep=2pt,
    arc=4pt,
    left=2pt,
    right=2pt,
    top=2pt,
    bottom=2pt
}
\usepackage{fancyhdr}
\usepackage{amsmath, amssymb}
\usepackage{newtxtext,newtxmath}

\fancypagestyle{plain}{
  \fancyhf{}
  \fancyhead[LE]{\thepage}
  
}
\title{AI Data Development: A Scorecard for the System Card Framework}
\author{
    Tadesse K. Bahiru, Haileleol Tibebu, and Ioannis A. Kakadiaris
}
\date{
    Computational Biomedicine Lab, Department of Computer Science, University of Houston, Houston, TX, USA \\
    \texttt{\{tbahiru, htibebu, ioannisk\}@uh.edu}
}

\begin{document}
\maketitle        

\begin{abstract}       
Artificial intelligence has transformed numerous industries, from healthcare to finance, enhancing decision-making through automated systems. However, the reliability of these systems is mainly dependent on the quality of the underlying datasets, raising ongoing concerns about transparency, accountability, and potential biases. This paper introduces a scorecard designed to evaluate the development of AI datasets, focusing on five key areas from the system card framework data development life cycle: data dictionary, collection process, composition, motivation, and pre-processing. The method follows a structured approach, using an intake form and scoring criteria to assess the quality and completeness of the data set. Applied to four diverse datasets, the methodology reveals strengths and improvement areas. The results are compiled using a scoring system that provides tailored recommendations to enhance the transparency and integrity of the data set. The scorecard addresses technical and ethical aspects, offering a holistic evaluation of data practices. This approach aims to improve the quality of the data set. It offers practical guidance to curators and researchers in developing responsible AI systems, ensuring fairness and accountability in decision support systems.\\

\end{abstract}

\section{Introduction}
Artificial intelligence technologies have become integral to various industries, significantly enhancing the efficiency and effectiveness of decision support systems in domains such as healthcare \cite{topol2019high}, finance \cite{cao2022ai}, criminal justice \cite{dressel2018accuracy}, and education \cite{zawacki2019systematic}. These systems promise increased accuracy and efficiency, but they also introduce critical concerns about transparency \cite{ananny_seeing_2018}, accountability \cite{diakopoulos_algorithmic_2015}, and biases \cite{selbst_fairness_2019}. As AI applications continue to expand \cite{haenlein2019brief}, it is essential to rigorously examine the datasets used to train and validate these systems. High-quality datasets are the backbone of reliable AI models \cite{shankar2017no}, ensuring these systems can make accurate and fair decisions. Without proper evaluation, these datasets may harbor hidden biases, such as those found in gender and skin type biases in facial recognition systems, which can undermine the integrity of AI-driven outcomes by causing disparities in performance across demographic groups \cite{buolamwini_gender_2018}.

The quality and integrity of datasets play a pivotal role in ensuring the reliability and fairness of decision support systems \cite{jagadish2019responsibility}. Biased data, incomplete documentation, and inadequate preprocessing can lead to significant ethical and operational challenges. For instance, biased datasets can perpetuate and even worsen societal inequities, as evidenced by studies highlighting algorithmic biases in decision support systems\cite{chouldechova_fair_2017}. On the other hand, well-documented datasets provide comprehensive insights into the data's structure, collection, and processing, making it easier to detect potential biases, imbalances, or other flaws that could affect the fairness of AI systems. This transparency is essential for ensuring that models are built on ethically sound and fair data foundations. As datasets grow in complexity and scale, they demand more sophisticated evaluation techniques to uncover and address hidden biases and inaccuracies \cite{koh2021wilds}. Recognizing these challenges, the AI community has emphasized the importance of detailed dataset documentation and standardization. Initiatives such as datasheets for datasets \cite{datasheet_data} and model cards \cite{mitchell_model_2019} have proposed standardized templates for documenting critical information about datasets and models. These efforts enhance transparency and accountability in AI development by providing detailed metadata about datasets' origins, composition, and intended use.

However, despite these advancements, there remains a gap in developing evaluation tools that assess the entire dataset life cycle. Existing approaches focus on specific aspects of dataset documentation without a complete view of critical areas such as data collection, composition, preprocessing, and overall quality \cite{datasheet_data}. The absence of such evaluations can lead to overlooked issues that affect the overall performance of Decision Support Systems. To address this gap, we adopt the system card approach \cite{gursoy2022system}, which assesses datasets based on critical data-related phases. Our scorecard framework is designed to serve both researchers by enhancing dataset transparency and reproducibility and industry practitioners by providing practical support for data governance and quality control. The evaluation focuses on five key assessment areas: data dictionary, data collection process, composition, motivation, and pre-processing. Evaluation criteria have been identified, each with a corresponding scoring rubric. Based on these criteria, an intake form with scoring guidelines has been developed for each area to ensure a clear and reliable approach to evaluating dataset development. The methodology is applied to four people-related datasets, highlighting areas where transparency in documentation can be improved. The evaluation results reveal both strengths and shortcomings of current dataset development practices, offering insights that contribute to refining data practices for better transparency and accountability. The main contributions of this paper are as follows: \begin{enumerate} 
\item We designed a methodology for evaluating data development using the System Card Framework, focusing on five key assessment areas.
\item We developed a scoring rubric for each area, enabling a more structured and consistent assessment of dataset development.
\item We introduced a scorecard summarizing the evaluation results and providing actionable recommendations to enhance dataset transparency and accountability. 
\end{enumerate}

\noindent The remainder of the paper is structured as follows: Section 2 reviews related work on dataset evaluation and documentation. Section 3 presents the proposed methodology for dataset evaluation. Section 4 details the evaluation results. Section 5 discusses the findings, and Section 6 concludes the paper with critical insights.
\section{Related Work}
With the rapid advancement of AI technology, there has been an increasing focus on evaluating and standardizing datasets to ensure the development of fair and accountable AI systems. One pioneering effort in this area is the Datasheets for Datasets initiative proposed by Gebru \textit{et al}. \cite{datasheet_data}. This work introduced a standardized template for documenting essential information about datasets, such as composition, collection process, motivation, pre-processing, and intended use. The authors argue that providing comprehensive dataset documentation can help mitigate biases, promote transparency, and facilitate responsible AI development. The Datasheets for Datasets initiative has been widely adopted and inspired similar field efforts.

Building upon the concept of dataset documentation, several researchers have proposed complementary frameworks. Holland  \textit{et al}. \cite{holland2020dataset} introduced a framework for dataset nutrition labels, which concisely summarize key dataset characteristics, such as provenance, quality, and potential biases. Bender and Friedman \cite{bender_data_2018} proposed the concept of data statements, designed to provide detailed information about the characteristics of natural language processing datasets. Mitchell  \textit{et al}. \cite{mitchell_model_2019} extended these ideas with model cards, documenting machine learning model performance, limitations, and intended use. Similarly, Bommasani \textit{et al.} \cite{bommasani2024fmtindex} introduced the Foundation Model Transparency Index (FMTI) to evaluate transparency across leading model developers using 100 indicators, such as data labor, compute, and model impact. The FMTI enhances accountability within the foundation model ecosystem by highlighting areas where transparency is limited, extending responsible AI principles from individual models to the entire development pipeline. Paullada  \textit{et al}. \cite{paullada2021data} emphasized the importance of comprehensive dataset documentation and standardized evaluation frameworks, highlighting the challenges of ensuring dataset quality and fairness, particularly with diverse and dynamic data sources. Arnold  \textit{et al}. \cite{arnold_factsheets_2019} proposed "Factsheets" to increase trust in AI services by documenting key details about AI systems, including performance, limitations, and safety, ensuring transparency and informed decision-making. Sambasivan  \textit{et al}. \cite{sambasivan_everyone_2021} highlighted the risks of "data cascades" in AI systems, where overlooked data quality issues lead to failures during deployment, emphasizing the importance of addressing data challenges early in the development life cycle. Similarly, Raji  \textit{et al}. \cite{raji2020closing} advocated for integrating dataset evaluations into AI auditing processes to enhance transparency and accountability, stressing that robust dataset documentation is critical to closing the accountability gap in AI systems. Gursoy  \textit{et al}. \cite{gursoy2022system} introduced the system card framework, which evaluates AI-based decision support systems through 56 criteria across four key areas: development, assessment, mitigation, and assurance. This approach ensures a holistic evaluation of AI systems, focusing on transparency, fairness, and accountability throughout their life cycle. Building on these efforts, the current research extends the system card framework to target the dataset life cycle, offering a more granular evaluation of key data-development components (Table \ref{tab:data_development}). Existing frameworks, such as Datasheets for Datasets and Model Cards, focus primarily on documentation but lack quantitative scoring mechanisms and comprehensive ethical evaluations. Our scorecard overcomes these limitations by integrating a scoring rubric, ethical transparency checks, and tailored recommendations, providing a holistic assessment of the dataset's quality. 
\vspace{-5pt}
\begin{table}[ht]
\centering
\renewcommand{\arraystretch}{1.2}
\caption{Basic Assessment Areas \cite{gursoy2022system}}\label{tab:data_development}
{\scriptsize
\begin{tabular}{p{3cm}p{6.5cm}p{1cm}}
\specialrule{1.2pt}{0pt}{0pt} 
\textbf{Category} & \textbf{Development Criteria} & \textbf{\#} \\ \hline
\multirow{5}{*}{Data} & Data dictionary & C111 \\ 
                     & Datasheet, Collection process & C112 \\ 
                     & Datasheet, Composition & C113 \\ 
                     & Datasheet, Motivation & C114 \\ 
                     & Datasheet, Preprocessing & C115 \\ 
\specialrule{1.2pt}{0pt}{0pt} 
\end{tabular}
}
\end{table}
\vspace{-5pt}
\section{Methodology}
The proposed methodology offers a structured approach to evaluating AI data-development practices, promoting transparency, accountability, and continuous improvement. This method focuses on five key assessment areas derived from the system card framework \cite{gursoy2022system}: data dictionary, collection process, composition, motivation, and pre-processing. For each area, intake forms and scoring rubrics were developed, which dataset owners completed, with responses collected and scored to provide recommendations for each area and an overall dataset scorecard. Scores for each criterion are averaged, resulting in a range from -1 to 1 for each assessment area. These scores are categorized into three color-coded levels, each reflecting a specific quality of data development practices. Scores below \( T_1 = 0.39 \) are labeled red, indicating significant deficiencies in data development characterized by a lack of consistency and thorough documentation. This level highlights areas requiring serious improvement in data documentation and development processes. Scores between \( T_1 = 0.39 \) and \( T_2 = 0.79 \) are labeled yellow, representing partial adherence to data development standards. Practices at this level demonstrate moderate quality but still require improvements in documentation, consistency, and overall data processes. Scores above \( T_2 = 0.79 \) are labeled green, reflecting exemplary data development practices where every step is well-documented, consistent, and reusable, showcasing adherence to best practices. These experimentally determined thresholds provide a framework to assess varying levels of data development quality and offer actionable insights for targeted improvements.
After calculating the scores and identifying the strengths and deficiencies of each assessment area using the criterion, a dataset development scorecard is produced to summarize findings across all assessment areas. Serving as a roadmap for dataset owners and providing end-users with a clear overview of strengths and weaknesses, it facilitates informed decision-making in model development projects. It promotes best practices in dataset development and transparency.
\subsection{Data Dictionary (C111)}\label{AA}
A data dictionary is a critical tool providing a structured metadata collection detailing a dataset's contents, format, and structure \cite{uhrowczik_data_1973}. The data dictionary evaluation intake form includes essential elements such as the dataset name, Digital Object Identifier (DOI), description, creation date, and version. 
\begin{table}[H]
\centering
\renewcommand{\arraystretch}{1.2} 
\caption{Data Dictionary (C111) Scoring Rubric}
\scriptsize
\begin{tabular}{p{3cm}p{6cm}p{3.8cm}}
\specialrule{1.2pt}{0pt}{0pt} 
\textbf{Evaluation Criterion} & \textbf{Description} & \textbf{Scoring Rubric} \\ \hline
\multicolumn{3}{l}{\textbf{Dataset Information}} \\
\hline
Dataset Name  & Indicates whether the official dataset name is provided in the data dictionary & \begin{tabular}[c]{@{}l@{}}- Provided (1)\\- Not provided (-1)\end{tabular} \\
\hline
Description & Describes whether a detailed dataset description is included in the data dictionary & \begin{tabular}[c]{@{}l@{}}- Detailed description provided (1)\\- Limited description provided (0)\\- No description provided (-1)\end{tabular} \\
\hline
Dataset Provider & Availability of dataset provider's contact information (e.g., email) & \begin{tabular}[c]{@{}l@{}}- Contact information provided (1)\\- Not provided (-1)\end{tabular} \\
\hline
DOI & Indicates the dataset's availability of a Digital Object Identifier (DOI) & \begin{tabular}[c]{@{}l@{}}- DOI provided (1)\\- DOI not provided (-1)\end{tabular} \\
\hline
Created Date & Indicates if the dataset creation date is provided in the data dictionary & \begin{tabular}[c]{@{}l@{}}- Creation date provided (1)\\- Creation date unknown (-1)\end{tabular} \\
\hline
Version & Indicates if the version information of the dataset is provided & \begin{tabular}[c]{@{}l@{}}- Version provided (1)\\- Version unknown (-1)\end{tabular} \\
\hline
\multicolumn{3}{l}{\textbf{Dataset Files}} \\
\hline
File Name & Lists all file names in the data dictionary, particularly for datasets with multiple files & \begin{tabular}[c]{@{}l@{}}- All files listed (1)\\- Some files listed (0)\\- No files listed (-1)\end{tabular} \\
\hline
Description & Indicates if a detailed description is provided for each file & \begin{tabular}[c]{@{}l@{}}- Detailed description provided (1)\\- No description provided (-1)\end{tabular} \\
\hline
Source & Indicates if information about the source of the files is provided & \begin{tabular}[c]{@{}l@{}}- Source provided for all files (1)\\- Source provided for some files (0)\\- No source provided (-1)\end{tabular} \\
\hline
\multicolumn{3}{l}{\textbf{Dataset Attributes}} \\
\hline
Attribute Name & Indicates if all attribute names in each file are provided in the data dictionary & \begin{tabular}[c]{@{}l@{}}- All attribute names provided (1)\\- Some attribute names missing (0)\\- No attribute names provided (-1)\end{tabular} \\
\hline
Parent File Name & Indicates if the file containing the attributes is specified in the dataset & \begin{tabular}[c]{@{}l@{}}- Provided for all attributes (1)\\- Provided for some attributes (0)\\- Parent file not provided (-1)\end{tabular} \\
\hline
Description & Indicates if a description is provided for each attribute in the data dictionary & \begin{tabular}[c]{@{}l@{}}- Provided for all attributes (1)\\- Provided for some attributes (0)\\- No description provided (-1)\end{tabular} \\
\hline
Data Type & Indicates if data types for each attribute are listed in the data dictionary & \begin{tabular}[c]{@{}l@{}}- Provided for all attributes (1)\\- Provided for some attributes (0)\\- No data type provided (-1)\end{tabular} \\
\hline
Allowed Value & Indicates if allowed values for each attribute are listed in the data dictionary & \begin{tabular}[c]{@{}l@{}}- Provided for all attributes (1)\\- Provided for some attributes (0)\\- No allowed values provided (-1)\end{tabular} \\
\hline
Missing Value & Indicates if the mechanism for handling missing values is described & \begin{tabular}[c]{@{}l@{}}- Handling mechanism provided (1)\\- Not provided (-1)\end{tabular} \\
\hline
Example Value  & Indicates if example values are provided for each attribute & \begin{tabular}[c]{@{}l@{}}- Provided for all attributes (1)\\- Provided for some attributes (0)\\- No example values provided (-1)\end{tabular} \\
\hline
Format & Indicates if the recorded data format for each attribute is provided & \begin{tabular}[c]{@{}l@{}}- Provided for all attributes (1)\\- Provided for some attributes (0)\\- No format provided (-1)\end{tabular} \\
\specialrule{1.2pt}{0pt}{0pt} 
\end{tabular}
\label{tab:dataset_components}
\end{table}

It also includes information on the files within the dataset, specifying their names, descriptions, and sources. We selected these criteria to evaluate the datasets' foundational documentation and ensure that essential metadata is available for accurate interpretation and practical use. Furthermore, the form records detailed attributes for each file, covering aspects such as attribute names, parent file names, descriptions, data types, allowed values, missing values, formats, and example values. These criteria evaluate the completeness and clarity of attribute-level documentation, which is crucial for users to understand data structures, variable meanings, and potential limitations of the datasets. Table \ref{tab:dataset_components} details the specific evaluation criteria, descriptions, and scoring methods, ensuring a comprehensive and standardized assessment of the data dictionary's documentation. 
\subsection{Collection Process (C112)}\label{BB}
The collection process is integral to dataset development, significantly influencing the quality and ethical considerations of the dataset \cite{olteanu_social_2019}. A well-documented data collection process is crucial for minimizing biases, ensuring data integrity, and improving the reliability of AI systems \cite{datasheet_data}. We assessed data source diversity and reliability to confirm that datasets are representative and unbiased, enhancing their generalizability and accuracy. Proper documentation of collection methods is essential, as it strengthens reproducibility and credibility.
\begin{table}[ht]
\centering
\renewcommand{\arraystretch}{1.2}
\scriptsize
\caption{Datasheet, Collection Process (C112) Scoring Rubric}
\begin{tabular}{p{2.5cm}p{6.3cm}p{6.1cm}}
\specialrule{1.2pt}{0pt}{0pt} 
\textbf{Evaluation Criterion} & \textbf{Description} & \textbf{Scoring Rubric} \\
\hline
Dataset Source  & Evaluates the diversity and reliability of the data sources used in dataset development & \begin{tabular}[c]{@{}l@{}}- Diverse and reliable sources, well-documented (1) \\ - Limited diversity and moderate documentation (0) \\ - Not sure and poorly documented (-1)\end{tabular} \\
\hline
Collection Techniques  & Assesses the appropriateness, consistency, and documentation of data collection methods used & \begin{tabular}[c]{@{}l@{}}- Appropriate, well-documented, consistent techniques (1)\\- Adequate techniques with partial documentation (0)\\- Inappropriate or poorly documented techniques (-1)\end{tabular} \\
\hline
Sampling Strategy & Evaluates the suitability and documentation of the dataset's sampling strategy, if applicable & \begin{tabular}[c]{@{}l@{}}- Well-suited to the dataset's goals and documented (1)\\ - Partially suitable with documentation (0)\\ - Poorly documented sampling strategy (-1)\end{tabular} \\
\hline
Involvement & Participants in data collection and details on compensation & \begin{tabular}[c]{@{}l@{}}- Well-documented involvement and compensation (1)\\- Partial documentation (0)\\- Limited documentation (-1)\end{tabular} \\
\hline
Time & Documents and assesses the timeframe during which data was collected & \begin{tabular}[c]{@{}l@{}}- The time frame is documented and known (1)\\- Timeframe is vaguely documented (0)\\- The time frame is undocumented and unknown (-1)\end{tabular} \\
\hline
Notification & Whether individuals were informed about data collection & \begin{tabular}[c]{@{}l@{}}- Individuals notified (1)\\- Individuals not notified (-1)\end{tabular} \\
\hline
Informed Consent & Whether individuals consented to use the data & \begin{tabular}[c]{@{}l@{}}- Consent obtained (1)\\- Consent not obtained (-1)\end{tabular} \\
\hline
Consent Revocation & Whether there is a way to revoke consent after data collection & \begin{tabular}[c]{@{}l@{}}- Clear and accessible mechanism available (1)\\- Mechanism available but not easily accessible (0)\\- No mechanism to revoke consent (-1)\end{tabular} \\
\hline
Ethical Review & Whether the data collection process underwent ethical review & \begin{tabular}[c]{@{}l@{}}- Comprehensive ethical guidelines followed (1)\\- Some ethical guidelines followed (0)\\- No ethical guidelines followed (-1)\end{tabular} \\
\hline
Limitations & Documented constraints or biases during data collection & \begin{tabular}[c]{@{}l@{}}- Identified and comprehensively documented (1)\\- Identified but documentation limited (0)\\- No limitation identification method (-1)\end{tabular} \\
\hline
Feedback & Mechanisms for gathering feedback from data users or participants & \begin{tabular}[c]{@{}l@{}}- Active mechanism for regular feedback (1)\\- Occasional feedback incorporation (0)\\- No mechanism for feedback incorporation (-1)\end{tabular} \\
\specialrule{1.2pt}{0pt}{0pt} 
\end{tabular}
\label{tab:dataset_collection_process}
\end{table}

At the same time, the sampling strategy provides insights into population representation, helping to identify any biases that may affect downstream analyses. Participant involvement and compensation were selected for evaluation to uphold transparency and fair treatment, acknowledging individual contributions to the dataset. We also consider the time frame of data collection, which directly impacts its relevance and contextual accuracy. Ethical considerations are core to our framework; ensuring that individuals are notified about data collection, provide informed consent, and have the option to revoke consent reinforces autonomy and privacy, fostering trust in the dataset's use. The ethical review of the data collection process is also crucial, confirming compliance with necessary guidelines and safeguarding participant rights. Transparency about any limitations encountered during data collection, such as methodological constraints or biases, informs users and aids in accurate data interpretation. Lastly, we include feedback mechanisms for continuous dataset improvement, supporting engagement and quality enhancement over time. Each criterion reinforces a robust, ethically sound collection process, promoting the development of datasets that uphold the standards of responsible AI while fostering trustworthiness, accountability, and rigor in AI applications. Table \ref{tab:dataset_collection_process} presents the evaluation criteria within the intake form and its scoring rubric. 
\subsection{Composition (C113)}\label{CC}

\begin{table}[ht]
\centering
\renewcommand{\arraystretch}{1.2}
\scriptsize
\caption{Datasheet, Composition (C113) Scoring Rubric}
\vspace{-0.3em}
\begin{tabular}{p{2.5cm}p{6.5cm}p{5.5cm}}
\specialrule{1.2pt}{0pt}{0pt} 
\textbf{Evaluation Criterion} & \textbf{Description} & \textbf{Scoring Rubric} \\
\hline
Number of Instances  & Total number of instances within the dataset, whether known or unknown & \begin{tabular}[c]{@{}l@{}}- Exact instance count known (1)\\- Approximate count known (0)\\- Instance count unknown (-1)\end{tabular} \\
\hline
Relationship  & Articulation of relationships between instances (e.g., user movie ratings, social network links) & \begin{tabular}[c]{@{}l@{}}- Relationships fully defined (1)\\- Partially defined relationships (0)\\- Relationships not defined (-1)\end{tabular} \\
\hline
Type  & Assesses diversity and complexity of instances in the dataset & \begin{tabular}[c]{@{}l@{}}- Single-type instances (e.g., documents, photos, people) (1)\\- Multiple types of instances (e.g., movies and ratings, \\ people and interactions,
  nodes and edges) (0)\\- Unclear or unspecified instance types (-1)\end{tabular} \\
\hline
Presence of Label  & Evaluates whether instances in the dataset have labels & \begin{tabular}[c]{@{}l@{}}- All instances labeled (1)\\- Some instances labeled (0)\\- No instances labeled (-1)\end{tabular} \\
\hline
Dataset Structure  & Describes how the dataset is divided into subsets (e.g., training, validation, testing) & \begin{tabular}[c]{@{}l@{}}- Training, validation, and testing data (1)\\- Training and testing data (1)\\- No specified partition (0)\end{tabular} \\
\hline
Dependencies  & Reliance on external sources and the quality of these dependencies & \begin{tabular}[c]{@{}l@{}}- The dataset is self-contained (1)\\- Dataset links to external resources (1)\\- External resource dependencies unclear (-1)\end{tabular} \\
\hline
Missing Information & Presence and clarity of missing data within instances & \begin{tabular}[c]{@{}l@{}}- All instances are complete (1)\\- Present in some instances (-1)\\- Unclear if any information is missing (0)\end{tabular} \\
\hline
Data Quality & Presence of quality issues like errors, noise, or redundancies & \begin{tabular}[c]{@{}l@{}}- No known errors, noise, or redundancies (1)\\- Limited errors, noise, or redundancies (0)\\- High errors, noise, or redundancies (-1)\end{tabular} \\
\hline
Subpopulation & Evaluates the identification and de-identification of subpopulations within the dataset & \begin{tabular}[c]{@{}l@{}}- No identifiable subpopulations (1)\\- Subpopulations are fully anonymized (1)\\- Subpopulations are identified (-1)\end{tabular} \\
\hline
Individual Identification & Potential for identifying individuals and the level of anonymization applied & \begin{tabular}[c]{@{}l@{}}- No identifiable information present (1)\\- Anonymized and protected (1)\\- Individuals can be identified and unprotected (-1)\end{tabular} \\
\hline
Sensitivity  & Sensitivity of the dataset content and its impact, including masking & \begin{tabular}[c]{@{}l@{}}- No potentially sensitive content (1)\\- Sensitive content is present and masked (1)\\- Sensitive content is present and unmasked (-1)\end{tabular} \\
\hline
Confidentiality  & Presence and protection of confidential data within the dataset & \begin{tabular}[c]{@{}l@{}}- No confidential data (1)\\- Confidential data is present and protected (1)\\- Confidential data is present and unprotected (-1)\end{tabular} \\
\specialrule{1.2pt}{0pt}{0pt} 
\end{tabular}
\label{tab:dataset_evaluation_criteria}
\end{table}

Dataset composition is a fundamental aspect of developing robust and unbiased AI systems. The structure, diversity, and effectiveness of the representation of the dataset significantly influence the performance and fairness of the machine learning models trained on it \cite{jo_lessons_2020}. Evaluating data composition is essential to identify potential shortcomings, such as underrepresented groups or hidden biases, which could lead to skewed results in downstream applications \cite{wilson_good_2017}. To thoroughly assess the quality of composition, our criteria, as outlined in Table \ref{tab:dataset_evaluation_criteria}, include essential elements, such as the number of instances and the relationships between them, which help establish the scope and context of the data. We examine labeling to ensure that instances are appropriately annotated, enhancing model interpretability and application. Subsetting into training, validation, and testing sets is assessed to confirm the dataset's readiness for model development. In addition to the intake form filled out by the dataset owner, we independently analyze the dataset to assess its quality, checking for errors, noise, and redundancy. This approach ensures that quality issues are identified and mitigated to prevent distortions in results. The presence of dependencies on external sources is reviewed to understand how they could impact consistency and accessibility. We also examine subpopulation representation and individual identifiability, as these factors influence fairness and privacy. To protect privacy, we assess confidentiality and sensitivity, ensuring sensitive or confidential data is handled appropriately.
\subsection{Motivation (C114)}\label{DD}
The motivation behind dataset development is essential for understanding its purpose, intended applications, and potential impact. In Table \ref{tab:dataset_motivation_criteria}, we present criteria that ensure each dataset is clearly defined, well-justified, and suited for specific applications. A well-articulated purpose statement provides immediate insight into the dataset's focus, helping users understand its aims and intended value. Addressing specific research gaps highlights how the dataset contributes to advancing knowledge or addressing real-world challenges. We include criteria that outline intended use cases to guide users toward appropriate applications, reducing potential misuse and enhancing the dataset's utility. Transparency about the development team, supporting organizations, and funding sources enables users to assess the expertise and resources needed to shape the dataset. Evaluating potential impact underscores the dataset's anticipated benefits and broader implications. 
\begin{table}[ht]
\centering
\renewcommand{\arraystretch}{1.2}
\scriptsize
\caption{Datasheet, Motivation (C114) Scoring Rubric}
\begin{tabular}{p{2.5cm}p{5cm}p{4.40cm}}
\specialrule{1pt}{0pt}{0pt} 
\textbf{Evaluation Criterion} & \textbf{Description} & \textbf{Scoring Rubric} \\
\hline
Purpose Statement & Clear articulation of the dataset's intended purpose & 
\begin{tabular}[c]{@{}l@{}}
- Clearly stated purpose (1)\\
- Vaguely stated purpose (0)\\
- No stated purpose (-1)
\end{tabular} \\
\hline
Research Gap Addressed & Identification of the specific research gap or problem the dataset aims to address & 
\begin{tabular}[c]{@{}l@{}}
- Identified research gap (1)\\
- Vaguely identified research gap (0)\\
- No mention of research gap (-1)
\end{tabular} \\
\hline
Intended Use Cases & Description of specific use cases or applications for the dataset & 
\begin{tabular}[c]{@{}l@{}}
- Comprehensive list of use cases (1)\\
- Limited use cases mentioned (0)\\
- No use cases specified (-1)
\end{tabular} \\
\hline
Development Team & Information about the team or individuals responsible for dataset creation & 
\begin{tabular}[c]{@{}l@{}}
- Detailed team information provided (1)\\
- Limited team information (0)\\
- No team information (-1)
\end{tabular} \\
\hline
Backing Organization & Identification of the organization(s) supporting the dataset development & 
\begin{tabular}[c]{@{}l@{}}
- Clearly stated backing organization (1)\\
- Vague mention of organization (0)\\
- No organization mentioned (-1)
\end{tabular} \\
\hline
Funding Sources & Disclosure of funding sources, including specific grants if applicable & 
\begin{tabular}[c]{@{}l@{}}
- Detailed funding information (1)\\
- Partial funding information (0)\\
- No funding information (-1)
\end{tabular} \\
\hline
Potential Impact & Discussion of the potential impact or benefits of the dataset & 
\begin{tabular}[c]{@{}l@{}}
- Comprehensive impact analysis (1)\\
- Limited impact discussion (0)\\
- No mention of potential impact (-1)
\end{tabular} \\
\specialrule{1pt}{0pt}{0pt} 
\end{tabular}
\label{tab:dataset_motivation_criteria}
\end{table}

\subsection{Preprocessing(C115)}\label{EE}
Each criterion in our preprocessing evaluation (Table \ref{tab:dataset_pre-processing_criteria}) is designed to provide a thorough and transparent data preparation assessment, focusing on technical rigor and alignment with the dataset's intended use. The preprocessing status criterion identifies whether the dataset has undergone any transformations or remains in its original form, providing context for any modifications that might impact usability. Steps applied ensure each preprocessing technique, such as data cleaning or transformations, is well-documented, allowing users to replicate or assess the suitability of these methods for their analyses. The software availability criterion checks if the tools used are accessible to external users, enhancing transparency and enabling verification by others. Consistent preprocessing application is vital; inconsistencies can introduce biases or errors, so this criterion ensures that transformations are uniformly applied across all data instances. The alignment criterion emphasizes that preprocessing steps should suit the dataset's analytical goals, as inappropriate transformations could distort results. Assessing the impact on data quality ensures that preprocessing reduces noise or other issues without compromising valuable information.
In contrast, the impact on data size monitors any reduction that could limit analysis potential. Verification of preprocessing steps confirms that transformations are validated, reducing the risk of unnoticed errors or biases. Retaining raw data alongside processed versions enables traceability, supporting transparency and reproducibility. Lastly, documentation clarity ensures each preprocessing ƒstep is explained in detail, helping end-users understand how data has been transformed and allowing them to make informed decisions regarding the dataset's suitability for specific applications.
\begin{table}[ht]
\centering
\renewcommand{\arraystretch}{1.2}
\scriptsize
\caption{Datasheet, Preprocessing(C115) Scoring Rubric}
\begin{tabular}{p{2.5cm}p{5.5cm}p{6cm}}
\specialrule{1.2pt}{0pt}{0pt} 
\textbf{Evaluation Criterion} & \textbf{Description} & \textbf{Scoring Rubric} \\
\hline
Preprocessing Status & Indicates whether the dataset has undergone preprocessing or remains in its original form & \begin{tabular}[c]{@{}l@{}}- preprocessing has been applied (1)\\- The data remains in its original, unprocessed form (-1)\end{tabular} \\
\hline
Steps Applied & Describes the specific preprocessing techniques used and their thoroughness & \begin{tabular}[c]{@{}l@{}}- Extensive, well-documented preprocessing (1)\\- Limited documentation and techniques (0)\\- Minimal or poorly documented preprocessing (-1)\end{tabular} \\
\hline
Software Availability & Specifies whether the software used for preprocessing is publicly available or restricted & \begin{tabular}[c]{@{}l@{}}- The software is publicly available (1)\\- The software is available under restricted conditions (0)\\- The software is not available (-1)\end{tabular} \\
\hline
Consistency & Assesses whether preprocessing is consistently applied across all data instances & \begin{tabular}[c]{@{}l@{}}- Uniformly applied across all data instances (1)\\- Varies across different segments of the dataset (-1)\end{tabular} \\
\hline
Alignment & Evaluates how well preprocessing aligns with the goals of data analysis & \begin{tabular}[c]{@{}l@{}}- preprocessing fully aligns with analysis goals (1)\\- Partially aligned, with room for improvement (0)\end{tabular} \\
\hline
Impact on Data Quality & Measures the effect of preprocessing on the quality of the dataset & \begin{tabular}[c]{@{}l@{}}- Improved data quality significantly (1)\\- Slight improvement in data quality (0)\\- Change or degradation in data quality (-1)\end{tabular} \\
\hline
Impact on Data Size & Assesses the extent of data loss or reduction during preprocessing& \begin{tabular}[c]{@{}l@{}}- No significant data loss (1)\\- Some loss, but within acceptable limits (0)\\- Significant data reduction (-1)\end{tabular} \\
\hline
Verification & Describes the processes in place to verify the preprocessing steps & \begin{tabular}[c]{@{}l@{}}- Comprehensive verification processes (1)\\- Limited verification; limited preprocessing steps verified (0)\\- No verification mechanism is provided (-1)\end{tabular} \\
\hline
Retention & Indicates whether raw data is retained alongside preprocessed versions & \begin{tabular}[c]{@{}l@{}}- Raw and preprocessed data retained (1)\\- Only preprocessed data retained (-1)\end{tabular} \\
\hline
Documentation Clarity & Evaluates the clarity and comprehensiveness of documentation for preprocessing techniques & \begin{tabular}[c]{@{}l@{}}- Well-documented preprocessing techniques (1)\\- Partial documentation, covering limited techniques (0)\\- No preprocessing documentation (-1)\end{tabular} \\
\specialrule{1.2pt}{0pt}{0pt} 
\end{tabular}
\label{tab:dataset_pre-processing_criteria}
\end{table}

\subsection{Data Scorecard}\label{FF}
The data scorecard, presented in Table \ref{tab:scorecard}, represents the outcome of our evaluation methodology, providing a structured and insightful summary of findings across five key assessment areas. This tool captures the dataset's strengths, weaknesses, tailored recommendations, and overall development stage, offering dataset owners, researchers, and practitioners a clear guide to understanding and addressing the dataset's specific characteristics. Scores for each assessment area are derived from responses on the intake form, allowing us to highlight notable strengths and pinpoint areas needing improvement. Based on these scores, a summarized scorecard is created, offering a clear view of the dataset's readiness and potential for growth.
\begin{table}
\caption{Data scorecard}
\begin{tcolorbox}
    \vspace{5pt}
    \vspace{-5pt} 
    \scriptsize
    \begin{tabular}{p{0.25\textwidth}p{0.60\textwidth}}
        \textbf{Dataset Name:} & [Dataset Name] \\
    \end{tabular}
    \vspace{0pt}
    \begin{tabular}{p{0.25\textwidth}p{0.70\textwidth}}
        \textbf{Description:} & [Brief description of the dataset, including its purpose, content, version, and  key features]
    \end{tabular}
    \vspace{0pt}
    
    \hrulefill 
    
    \vspace{0pt}
    \begin{tabular}{p{0.25\textwidth}p{0.18\textwidth}p{0.50\textwidth}}
        \textbf{Criteria} & \textbf{Document Available?} & \multicolumn{1}{c}{\textbf{Score / Assessment Status}} \\
        \hline
        \textbf{Data Dictionary} & Yes/No & \textbf{[Score] [Status]} \newline \textbf{Remarks:} [Evaluate the clarity, detail, and comprehensiveness of the data dictionary. Note if all attributes, data types, and value ranges are well-documented.] \\
        \textbf{Collection Process} & Yes/No & \textbf{[Score] [Status]} \newline \textbf{Remarks:} [Assess the thoroughness of the data collection documentation, including sources, methods, and ethical considerations. Ensure it includes participant consent and data handling procedures.] \\
        \textbf{Composition} & Yes/No & \textbf{[Score] [Status]} \newline \textbf{Remarks:} [Examine the demographic breakdown, diversity, biases, and overall structure of the dataset. Comment on any imbalances or representational biases.] \\
        \textbf{Motivation} & Yes/No & \textbf{[Score] [Status]} \newline \textbf{Remarks:} [Analyze the stated purpose and intended use cases of the dataset. Ensure the motivations align with the dataset's structure and content.] \\
        \textbf{Prerocessing} & Yes/No & \textbf{[Score] [Status]} \newline \textbf{Remarks:} [Review the documentation of preprocessing steps, including methods, consistency, and impact on data quality.] \\
    \end{tabular}
    \vspace{0pt}
    
    \hrulefill 
    
    \vspace{0pt}
    
    \textbf{Overall Assessment:} [Summary of the strengths and weaknesses of the dataset based on the evaluation criteria. Highlight key areas where the dataset excels and where it needs improvement.]
    
    \vspace{0pt} 
    \hrulefill
    
    \vspace{0pt}
    \textbf{Recommendations:} [Specific recommendations for improving the dataset documentation and quality. Recommend actionable steps for the dataset owners to address the identified issues and enhance the overall utility.]
\end{tcolorbox}
\label{tab:scorecard}
\end{table}
The scorecard begins by providing essential information about the dataset, such as its name, owner, version, and a brief description, setting the necessary context for the evaluation. It then presents a detailed breakdown of the evaluation results for each assessment area, with each area assigned an overall score and a color-coded categorization (green, yellow, or red) that allows for a quick visual assessment of the dataset's performance. The color-coded system helps identify areas where the dataset performs well, improvements are needed, and critical issues may exist. In addition to the overall score, the scorecard summarizes the key findings within each assessment area, highlighting strengths. These pinpoint areas that need improvement and identify critical gaps or concerns uncovered during the evaluation process. The scorecard also includes tailored recommendations and specific action items for dataset owners, providing practical steps for addressing the identified shortcomings and guiding them toward continuous improvement. These suggestions help owners enhance the dataset's quality, ensuring it aligns with best practices and supports responsible, effective use in AI and research applications.

\section{Results}

The evaluation of the proposed methodology was conducted using four diverse datasets: Labeled Faces in the Wild (LFW) \cite{huang_labeled_2008}, MIMIC-IV \cite{johnson_mimic-iv_nodate}, NIJ's recidivism \cite{noauthor_nijs_nodate}, and the Baylor College of Medicine dataset (BCM-A) \cite{bautista2023ai}. These datasets were selected for their significance, data quality, complexity, and available metadata, representing face recognition, criminal justice, and health records benchmarks. LFW is a widely used benchmark for face recognition, containing labeled web-gathered face photographs that are instrumental in advancing facial recognition technology. MIMIC-IV, an extensive publicly available database, includes de-identified health data from critical care patients, making it valuable for healthcare predictive modeling. The recidivism dataset tracks post-incarceration patterns of approximately 26,000 individuals, providing critical insights for criminal justice reform and analysis. The BCM-A dataset is a comprehensive collection of electronic health records aimed at developing predictive models for conditions like MIS-C and murine typhus. The evaluation results, including average scores across the five key assessment areas, are summarized in Table \ref{tab:datasets_aspects}, with sample scorecards for LFW and MIMIC-IV in Tables \ref{tab:lfw-scorecard} and \ref{tab:mimic-scorecard} demonstrating the methodology's application. As described in the methodology section, an intake form with a dropdown list was prepared and provided to each dataset's owner for completion, allowing us to collect relevant information on dataset development. While the dropdown list did not display scores to the dataset owners during completion, the scores became visible upon submission. Once the forms were filled and submitted, we calculated scores based on the scoring rubric given for each assessment area, identifying both the strengths and weaknesses of the dataset development. Based on the scores in each assessment area, a final scorecard was created to provide an in-depth evaluation. 

\begin{table}[ht]
    \centering
    \setlength{\tabcolsep}{1pt} 
    \caption{Evaluation result of the selected datasets}
    \scriptsize
    \begin{tabular}{p{1.5cm}p{1cm}p{0.90cm}p{1cm}p{1cm}p{1cm}p{1cm}p{1cm}p{1cm}p{0.90cm}p{0.90cm}}
        \specialrule{1.2pt}{0pt}{0pt} 
        & \multicolumn{2}{p{1.5cm}}{\textbf{\centering Data Dictionary}} 
        & \multicolumn{2}{p{1cm}}{\textbf{\centering Collection}} 
        & \multicolumn{2}{p{1cm}}{\textbf{\centering Composition}} 
        & \multicolumn{2}{p{1cm}}{\textbf{\centering Motivation}} 
        & \multicolumn{2}{p{1cm}}{\textbf{\centering Preprocessing}} \\
        \hline
        \textbf{Dataset} & \textbf{Score} & \textbf{Color} 
        & \textbf{Score} & \textbf{Color} 
        & \textbf{Score} & \textbf{Color} 
        & \textbf{Score} & \textbf{Color} 
        & \textbf{Score} & \textbf{Color} \\
        \hline
        LFW & -1.00 & Red & 0.17 & Red & 0.77 & Yellow & 1.00 & Green & 0.80 & Green \\
        MIMIC-IV & 0.92 & Green & 0.33 & Red & 0.85 & Green & 1.00 & Green & 0.70 & Yellow \\
        Recidivism & 1.00 & Green & -0.08 & Red & 0.85 & Green & 1.00 & Green & -1.00 & Red \\
        BCM-A & 1.00 & Green & 0.08 & Red & 0.92 & Green & 1.00 & Green & -1.00 & Red \\
        \specialrule{1.2pt}{0pt}{0pt} 
    \end{tabular}
    \label{tab:datasets_aspects}
\end{table}

\textit{Data Dictionary}: The data dictionary evaluation reveals notable differences among the datasets. The LFW dataset scored poorly, with a -1.00 rating and a red color, primarily due to the absence of a comprehensive data dictionary. This deficiency significantly hinders understanding the data structure and semantics, making the dataset challenging for more complex tasks. On the other hand, the MIMIC-IV dataset performed better, achieving a score of 0.92 and a green rating. While all necessary information is available on its website, unlike other datasets, the MIMIC-IV dataset lacks a standalone, well-organized data dictionary document. Nevertheless, the online resources provided most of the required details. The Recidivism and BCM-A datasets performed exceptionally well in this category, earning perfect scores of 1.00 with green color ratings. These datasets offered comprehensive data dictionaries, covering file-level and attribute-level details, ensuring clarity and ease of use.\\
\indent \textit{Collection Process}: The LFW dataset exhibited significant deficiencies, receiving a score of -1.00 and a red rating. It lacked critical elements, including notifications of data collection, informed consent, consent revocation mechanisms, and ethical review processes. This absence makes it difficult to assess its data collection methods' appropriateness and ethical considerations. The MIMIC-IV dataset performed better, scoring 0.33. While it adhered to ethical guidelines through data anonymization, it lacked sufficient documentation on consent revocation, feedback mechanisms, and limitations during the data collection. The Recidivism dataset scored -0.08 and revealed similar gaps in its documentation of informed consent, consent revocation, and ethical review processes, leaving room for improvement. Lastly, the BCM-A dataset scored 0.08, providing some documentation on data sources, collection mechanisms, and sampling strategies but missing key details on personnel involvement, informed consent, and feedback mechanisms. Overall, LFW exhibited the most significant gaps, while MIMIC-IV showed better adherence to ethical guidelines and documentation completeness.\\
\indent \textit{Composition}: The LFW dataset exhibited notable shortcomings in its composition, such as not articulating relationships among instances, potential identification of individuals, and the presence of sensitive information. The MIMIC-IV dataset faced similar issues, including missing and noisy data. The Recidivism and BCM-A datasets had good composition overall but included missing data, noise, and errors. Despite these issues, other criteria such as the number of instances, completeness, presence of labels, structure, and handling of missing data information were well-documented across all datasets except LFW.\\
\indent \textit{Motivation}: All datasets demonstrated strong motivation, with clearly stated purposes, identified research gaps, and comprehensive lists of intended use cases. Detailed information about development teams, backing organizations, and funding sources was provided. Potential impacts were well-documented, discussing how these datasets could improve predictive models and aid policy formulation. This thorough documentation led to perfect scores for the motivation criteria across all datasets.\\
\indent \textit{Preprocessing}: The LFW dataset met key preprocessing criteria, such as consistency, alignment, and improvements in data quality, but lacked clear documentation on verification processes and their impact on usability. Despite these gaps, it scored 0.80 with a green rating. The MIMIC-IV dataset demonstrated some preprocessing steps, including data cleaning and transformation, but lacked a well-structured document outlining these processes. While it provided partial pre-processing, further steps are necessary for full usability. The Recidivism dataset had minimal documentation, showing significant deficiencies in consistency, verification, and data retention, resulting in a low score and limiting utility. The BCM-A dataset was provided in its raw form, without pre-processing, making it ineligible for evaluation.
\begin{table}
\caption{Labeled Faces in the Wild dataset scorecard}
\begin{tcolorbox}
    \vspace{3pt}
    \vspace{-5pt} 
    \scriptsize
    \begin{tabular}{p{0.25\textwidth}p{0.6\textwidth}}
        \textbf{Dataset Name:} & Labeled Faces in the Wild \\
    \end{tabular}
    \vspace{3pt}
    \begin{tabular}{p{0.25\textwidth}p{0.69\textwidth}}
        \textbf{Description:} & A large-scale dataset containing face photographs designed to support the study of unconstrained face recognition. It comprises over 13,000 images of faces collected from the web, with pose, lighting, and expression variations. \\
    \end{tabular}
    \vspace{0pt}
    
    \hrulefill 
    
    \vspace{0pt}
    \begin{tabular}{p{0.25\textwidth}p{0.18\textwidth}p{0.50\textwidth}}
        \textbf{Criteria} & \textbf{Document Available?} & \multicolumn{1}{c}{\textbf{Score / Assessment Status}} \\
        \hline
        \textbf{Data Dictionary} & No & \textbf{[-1.0] [Red]} \newline \textbf{Remarks:} The dataset lacks a data dictionary document, hindering the understanding of its structure and semantics. Including detailed descriptions of attributes and their values is necessary. \\
        \textbf{Collection Process} & Yes & \textbf{[0.17] [Red]} \newline \textbf{Remarks:} The collection process is insufficiently documented, with missing details on consent, ethical reviews, and participant notifications. Proper docu documentation is essential for transparency. \\
        \textbf{Composition} & Yes & \textbf{[0.77] [Yellow]} \newline \textbf{Remarks:} The dataset has a diverse set of instances, but there are concerns about data quality, such as errors and noise. Ensuring data consistency and accuracy is crucial. \\
        \textbf{Motivation} & Yes & \textbf{[1.0] [Green]} \newline \textbf{Remarks:} The purpose and use cases are well-defined, focusing on addressing specific research gaps in face recognition. The dataset's relevance to current research is clearly articulated. \\
        \textbf{pre-processing} & Yes & \textbf{[0.80] [Green]} \newline \textbf{Remarks:} The preprocessing steps are well-documented and have contributed positively to data quality. Ensuring consistency across all instances enhances the dataset's utility. \\
    \end{tabular}
    \vspace{0pt}
    \hrulefill 
    \vspace{0pt}
    
    \textbf{Overall Assessment:} The dataset demonstrates strong motivation and good composition. However, it significantly lacks documentation, including a data dictionary and collection process details, affecting usability and transparency.

    \vspace{0pt} 
    \hrulefill
    
    \vspace{0pt}
    \textbf{Recommendations:}
\begin{itemize}
    \item \textbf{Data Dictionary:} Create and include a comprehensive data dictionary that provides detailed descriptions of each attribute, including data types, possible values, and examples. This will greatly enhance researchers' understanding of the dataset's usability.
    \item \textbf{Collection Process:} Significantly improve the documentation of the data collection process by including thorough details on how the data was collected, including participant consent, ethical review procedures, and notifications. This is critical for ensuring the transparency and ethical use of the dataset.
    \item \textbf{Composition:} Focus on improving data quality by addressing identified errors and noise. Implement regular data validation and cleaning processes to ensure the dataset's integrity and consistency.
    \item \textbf{Motivation:} Maintain the strong articulation of the dataset's purpose and relevance to face recognition research. Continue to update the motivation section as new research gaps or applications emerge to align the dataset with current needs.
    \item \textbf{pre-processing:} Ensure that the preprocessing steps are consistently applied across all instances in the dataset. Additionally, provide clear and detailed documentation of the preprocessing methods used, including any tools or algorithms applied. Regularly verify preprocessing outcomes to maintain high data quality.
\end{itemize}

\end{tcolorbox}
\label{tab:lfw-scorecard}
\end{table}

\begin{table}[H]
\caption{MIMIC-IV dataset scorecard}
\begin{tcolorbox}
    \vspace{3pt}
    \vspace{-5pt} 
    \scriptsize
    \begin{tabular}{p{0.25\textwidth}p{0.6\textwidth}}
        \textbf{Dataset Name:} & MIMIC-IV \\
    \end{tabular}
    \vspace{3pt}
    \begin{tabular}{p{0.25\textwidth}p{0.69\textwidth}}
        \textbf{Description:} & A large, freely-available database comprising de-identified health-related data associated with over 40,000 critical care patients, providing detailed information about patient demographics, vital signs, laboratory tests, medications, and outcomes. \\
    \end{tabular}
    \vspace{0pt}
    
    \hrulefill 
    
    \vspace{0pt}
    \begin{tabular}{p{0.25\textwidth}p{0.18\textwidth}p{0.50\textwidth}}
        \textbf{Criteria} & \textbf{Document Available?} & \multicolumn{1}{c}{\textbf{Score / Assessment Status}} \\
        \hline
        \textbf{Data Dictionary} & Yes & \textbf{[0.82] [Green]} \newline \textbf{Remarks:} Comprehensive and clear data dictionary, including detailed descriptions of attributes and data types. Further enhancement with more specific attribute-level details would be beneficial. \\
        \textbf{Collection Process} & Yes & \textbf{[0.33] [Red]} \newline \textbf{Remarks:} The documentation lacks detailed information on consent procedures, ethical reviews, and participant notifications. Improving these areas is essential for full transparency. \\
        \textbf{Composition} & Yes & \textbf{[0.85] [Green]} \newline \textbf{Remarks:} The dataset is generally comprehensive with a diverse set of instances, but there are some concerns about data quality, such as errors and noise. Ensuring consistent data accuracy is crucial. \\
        \textbf{Motivation} & Yes & \textbf{[1.0] [Green]} \newline \textbf{Remarks:} The purpose and intended use cases are articulated, with a comprehensive analysis of the potential impact, addressing research gaps effectively. \\
        \textbf{pre-processing} & Yes & \textbf{[0.70][Yellow]} \newline \textbf{Remarks:} The dataset demonstrated some preprocessing steps, including data cleaning and transformation. However, the documentation is poorly structured and lacks a clear outline of these processes.\\
    \end{tabular}
    \vspace{0pt}
    \hrulefill 
    \vspace{0pt}
    
    \textbf{Overall Assessment:}   The dataset is well-documented and has a clear purpose. However, it presents some data quality issues and lacks detailed information on data collection procedures, especially concerning ethical considerations and consent. Additionally, the preprocessing steps are not thoroughly documented and need more clarity.

    \vspace{0pt} 
    \hrulefill
    
    \vspace{0pt}

\textbf{Recommendations:}
\begin{itemize}
    \item \textbf{Data Dictionary:} Expand the data dictionary to include more granular attribute-level details, such as specific examples of values for each attribute and clarification of any ambiguous terms.
    \item \textbf{Collection Process:} Significantly improve the documentation of the data collection process by including detailed information on informed consent procedures, ethical reviews, and participant notifications. This should include explicit details on how consent was obtained, the scope of ethical reviews conducted, and how participants were informed about the data collection.
    \item \textbf{Composition:} Although the dataset is diverse and well-structured, focus on enhancing data quality by addressing existing errors and noise. Implement systematic data cleaning procedures and document these efforts to ensure consistent data accuracy.
    \item \textbf{Motivation:} Maintain the strong articulation of the dataset's purpose and impact. However, consider updating the motivation section regularly to reflect any new research gaps or emerging use cases as they arise.
    \item \textbf{Pre-processing:} Improve the clarity and detail of preprocessing documentation, ensuring that all steps are comprehensively described. Additionally, consider implementing a more robust verification process for preprocessing steps to ensure consistency across the dataset. Providing access to raw and preprocessed data for transparency is also recommended.
\end{itemize}

\end{tcolorbox}
\label{tab:mimic-scorecard}
\end{table}

\section{Discussion}
Evaluating the selected datasets using our proposed methodology revealed significant documentation quality and completeness disparities, which are essential for ethical and practical model development. Deficiencies in ethical transparency, such as missing consent documentation or lack of ethical review processes, can lead to AI models trained on biased or non-compliant data, perpetuating unfair outcomes, reducing model reliability, and eroding user trust. For instance, datasets without proper consent mechanisms may introduce privacy risks, while insufficient documentation of preprocessing steps can lead to inconsistencies that degrade model performance. Addressing these deficiencies is crucial for developing fair, accountable, and trustworthy AI systems. Our findings highlight several critical gaps in existing dataset evaluation frameworks, such as Datasheets for Datasets and Model Cards, which focus primarily on documentation without offering comprehensive life cycle evaluations or quantitative assessments. The scorecard methodology addresses these gaps by identifying documentation deficiencies and providing structured, actionable insights. For example, the lack of ethical transparency and preprocessing consistency observed in LFW and MIMIC-IV demonstrates the need for a more holistic evaluation approach. The scorecard's ability to pinpoint these deficiencies and suggest improvements fills a significant void in current practices, enabling dataset curators to enhance transparency, integrity, and fairness.

LFW dataset exhibited substantial deficiencies, particularly in its lack of a comprehensive data dictionary, which severely limits understanding its structure and contents. Furthermore, the data collection process documentation was inadequate, with minimal clarity around participant consent and ethical considerations, raising serious concerns about the dataset's usability, data quality, and ethical transparency. Similarly, the MIMIC-IV dataset, though performing better in motivation and composition with a clear purpose and well-structured data, still lacked sufficient documentation in areas such as the data dictionary and collection process. The absence of a standalone data dictionary, incomplete documentation of participant consent, and limitations reduced the dataset's transparency. Additionally, while some preprocessing steps were documented in MIMIC-IV, further clarification is required to ensure the data is fully ready for complex analyses. The gaps indicate that both LFW and MIMIC-IV require improvements in documentation to enhance their usability and ensure their ethical integrity. The Recidivism and BCM-A datasets performed better in documenting key attributes such as their data dictionaries and overall composition, which aids in understanding the datasets and supports their usability. However, both datasets revealed substantial shortcomings in documenting the data collection process, particularly regarding participant consent, ethical transparency, and participant involvement. These gaps raise concerns about the datasets' trustworthiness, as the lack of clear documentation around ethical considerations could impact the fairness and reliability of models developed using these data. preprocessing documentation was also a consistent area of concern. While providing some preprocessing information, the MIMIC-IV dataset lacked thorough verification and consistency, leaving room for improvement. The Recidivism dataset provided minimal documentation of its preprocessing steps, limiting more robust readiness for model development. The BCM-A dataset, provided in its raw form without any pre-processing, could not be evaluated in this regard, leaving a critical gap in understanding its suitability. LFW, too, failed to document preprocessing steps adequately, meeting only essential criteria, which impacts the overall trust in its data quality. These findings emphasize the necessity for more robust documentation practices, including more detailed data dictionaries, better transparency in data collection processes, and more transparent, more comprehensive preprocessing documentation.
\section{Conclusion}
The proposed methodology for evaluating data development effectively addresses critical gaps in existing frameworks, offering a thorough and structured approach for assessing dataset development. Its application to the selected datasets demonstrated the method's capacity to identify critical issues in areas such as the data dictionary, data collection process, composition, motivation, and pre-processing. This methodology underscores the importance of thorough documentation in ensuring dataset usability, reliability, and ethical transparency. Including the data-development scorecard enhances usability by concisely summarizing evaluation results and offering actionable insights for dataset owners. Addressing ethical transparency, incomplete documentation, and preprocessing inconsistencies mitigates risks associated with biased and unreliable datasets. These improvements ensure that AI systems are built on ethically sound and transparent data foundations. The evaluation framework supports dataset curators in refining their practices and helps model developers and researchers adhere to fairness, accountability, and ethical AI principles. The findings contribute to the broader goal of responsible AI development, where high-quality datasets foster trust, reduce biases, and enhance the integrity of automated decision-making systems. Future work will focus on integrating automated scoring tools to streamline evaluations and analyze sensitive, confidential data and re-identification risks, which are not fully addressed in this paper.
\section*{Acknowledgment}
 We thank our colleagues, Professors Ryan Kennedy, Andrew C. Michaels, and Lydia B. Tied, for their valuable support. We are also grateful to the CRASA project community advisors team, including Michael O. Adams, Katherine A. Franco, Sharon A. Israel, Jay Jenkins, James Marcella, Paula C. Pineda, and Jani Maselli Wood.

The National Science Foundation, under award number 2131504, partly supported this research. The views and conclusions expressed in this paper are those of the authors and do not necessarily reflect the official policies or endorsements, either expressed or implied, of the National Science Foundation.

\bibliographystyle{unsrt}

\begin{thebibliography}{10}

\bibitem{topol2019high}
E.~J. Topol.
\newblock High-performance medicine: the convergence of human and artificial intelligence.
\newblock {\em Nature Medicine}, 25(1):44--56, (2019).

\bibitem{cao2022ai}
L.~Cao.
\newblock {AI} in finance: challenges, techniques, and opportunities.
\newblock {\em ACM Computing Surveys}, 55(3):1--38, (2022).

\bibitem{dressel2018accuracy}
J.~Dressel and H.~Farid.
\newblock The accuracy, fairness, and limits of predicting recidivism.
\newblock {\em Science Advances}, 4(1), (2018).

\bibitem{zawacki2019systematic}
O.~Z.~Richter, V.~I Mar{\'\i}n, M.~Bond, and F.~Gouverneur.
\newblock Systematic review of research on artificial intelligence applications in higher education--where are the educators?
\newblock {\em International Journal of Educational Technology in Higher Education}, 16(1):1--27, (2019).

\bibitem{ananny_seeing_2018}
M.~Ananny and K.~Crawford.
\newblock Seeing without knowing: Limitations of the transparency ideal and its application to algorithmic accountability.
\newblock {\em New Media \& Society}, 20(3):973--989, (2018).

\bibitem{diakopoulos_algorithmic_2015}
N.~Diakopoulos.
\newblock Algorithmic accountability: Journalistic investigation of computational power structures.
\newblock {\em Digital Journalism}, 3(3):398--415, (2015).

\bibitem{selbst_fairness_2019}
A.~D. Selbst, Danah Boyd, S.~A. Friedler, S.~Venkatasubramanian, and J.~Vertesi.
\newblock Fairness and abstraction in sociotechnical systems.
\newblock In {\em Proc. Conference on Fairness, Accountability, and Transparency}, pages 59--68, Atlanta, GA, January 29--31, (2019).

\bibitem{haenlein2019brief}
M.~Haenlein and A.~Kaplan.
\newblock A brief history of artificial intelligence: On the past, present, and future of artificial intelligence.
\newblock {\em California Management Review}, 61(4):5--14, (2019).

\bibitem{shankar2017no}
S.~Shankar, Y.~Halpern, E.~Breck, J.~Atwood, J.~Wilson, and D.~Sculley.
\newblock No classification without representation: Assessing geodiversity issues in open data sets for the developing world.
\newblock {\em arXiv preprint arXiv:1711.08536}, (2017).

\bibitem{buolamwini_gender_2018}
J.~Buolamwini and T.~Gebru.
\newblock Gender shades: Intersectional accuracy disparities in commercial gender classification.
\newblock In {\em Proc. Conference on Fairness, Accountability and Transparency}, pages 77--91, New York, NY, January 21--24, (2018).

\bibitem{jagadish2019responsibility}
H.V. Jagadish, F.~Bonchi, T.~E.~Rad, L.~Getoor, K.~Gummadi, and J.~Stoyanovich.
\newblock The responsibility challenge for data.
\newblock In {\em Proc. Conference on Management of Data}, pages 412--414, Amsterdam, Netherlands, July 1--4, (2019).

\bibitem{chouldechova_fair_2017}
A.~Chouldechova.
\newblock Fair prediction with disparate impact: A study of bias in recidivism prediction instruments.
\newblock {\em Big Data}, 5(2):153--163, (2017).

\bibitem{koh2021wilds}
P.~W. Koh, S.~Sagawa, H.~Marklund, S.~M. Xie, M.~Zhang, A.~Balsubramani, W.~Hu, M.~Yasunaga, R.~L. Phillips, I.~Gao, T.~Lee, E.~David, I.~Stavness, W.~Guo, B.~Earnshaw, I.~Haque, S.~M. Beery, J.~Leskovec, A.~Kundaje, E.~Pierson, S.~Levine, C.~Finn, and P.~Liang.
\newblock {WILDS}: A benchmark of in-the-wild distribution shifts.
\newblock In {\em Proc. the 38th International Conference on Machine Learning}, pages 5637--5664, July 18–24, (2021).

\bibitem{datasheet_data}
T.~Gebru, J.~Morgenstern, B.~Vecchione, J.~W. Vaughan, H.~Wallach, H.~D.~III, and K.~Crawford.
\newblock Datasheets for datasets.
\newblock {\em Communications of the {ACM}}, 64(12):86--92, (2021).

\bibitem{mitchell_model_2019}
M.~Mitchell, S.~Wu, A.~Zaldivar, P.~Barnes, L.~Vasserman, B.~Hutchinson, E.~Spitzer, I.~D. Raji, and T.~Gebru.
\newblock Model cards for model reporting.
\newblock In {\em Proc. Conference on Fairness, Accountability, and Transparency}, pages 220--229, Atlanta, GA, January 29--31, (2019).

\bibitem{gursoy2022system}
F.~Gursoy and I.~A. Kakadiaris.
\newblock System cards for {AI}-based decision-making for public policy.
\newblock {\em arXiv preprint arXiv:2203.04754}, (2022).

\bibitem{holland2020dataset}
S.~Holland, A.~Hosny, S.~Newman, J.~Joseph, and K.~Ch.
\newblock The dataset nutrition label.
\newblock {\em Data Protection and Privacy}, 12(12):1, (2020).

\bibitem{bender_data_2018}
E.~M. Bender and B.~Friedman.
\newblock Data statements for natural language processing: Toward mitigating system bias and enabling better science.
\newblock {\em Transactions of the Association for Computational Linguistics}, 6:587--604, (2018).

\bibitem{bommasani2024fmtindex}
R.~Bommasani, K.~Klyman, S.~Longpre, S.~Kapoor, N.~Maslej, B.~Xiong, D.~Zhang, and P.~Liang.
\newblock The foundation model transparency index.
\newblock {\em arXiv preprint arXiv:2310.12941}, (2023).

\bibitem{paullada2021data}
A.~Paullada, I.~D. Raji, E.~M Bender, E.~Denton, and A.~Hanna.
\newblock Data and its (dis) contents: A survey of dataset development and use in machine learning research.
\newblock {\em Patterns}, 2(11), (2021).

\bibitem{arnold_factsheets_2019}
M.~Arnold, R.~K.~E. Bellamy, M.~Hind, S.~Houde, S.~Mehta, A.~Mojsilovic, R.~Nair, K.~Natesan Ramamurthy, A.~Olteanu, D.~Piorkowski, D.~Reimer, J.~Richards, J.~Tsay, and K.~R. Varshney.
\newblock {FactSheets}: Increasing trust in {AI} services through supplier's declarations of conformity.
\newblock {\em {IBM} Journal of Research and Development}, 63(4):6:1--6:13, (2019).

\bibitem{sambasivan_everyone_2021}
N.~Sambasivan, S.~Kapania, H.~Highfill, D.~Akrong, P.~Paritosh, and L.~M. Aroyo.
\newblock {“E}veryone wants to do the model work, not the data work”: Data cascades in high-stakes {AI}.
\newblock In {\em Proc. CHI Conference on Human Factors in Computing Systems}, pages 1--15, Yokohama, Japan, May 8--13, (2021).

\bibitem{raji2020closing}
I.~D. Raji, A.~Smart, R.~N. White, M.~Mitchell, T.~Gebru, B.~Hutchinson, J.~S.~Loud, D.~Theron, and P.~Barnes.
\newblock Closing the {AI} accountability gap: Defining an end-to-end framework for internal algorithmic auditing.
\newblock In {\em Proc. Conference on Fairness, Accountability, and Transparency}, pages 33--44, New York, NY, January 27--30, (2020).

\bibitem{uhrowczik_data_1973}
P.~P. Uhrowczik.
\newblock Data dictionary/directories.
\newblock {\em {IBM} Systems Journal}, 12(4):332--350, (1973).

\bibitem{olteanu_social_2019}
A.~Olteanu, C.~Castillo, F.~Diaz, and E.~Kıcıman.
\newblock Social data: Biases, methodological pitfalls, and ethical boundaries.
\newblock {\em Frontiers in Big Data}, 2:13, (2016).

\bibitem{jo_lessons_2020}
E.~S. Jo and T.~Gebru.
\newblock Lessons from archives: Strategies for collecting sociocultural data in machine learning.
\newblock In {\em Proc. Conference on Fairness, Accountability, and Transparency}, pages 306--316, Barcelona, Spain, January 27--30, (2020).

\bibitem{wilson_good_2017}
G.~Wilson, J.~Bryan, K.~Cranston, J.~Kitzes, L.~Nederbragt, and T.~K. Teal.
\newblock Good enough practices in scientific computing.
\newblock {\em {PLOS} Computational Biology}, 13(6), (2017).

\bibitem{huang_labeled_2008}
G.~B. Huang, M.~Mattar, T.~Berg, and E.~L.~Miller.
\newblock Labeled faces in the wild: A database for studying face recognition in unconstrained environments.
\newblock In {\em Proc. Workshop on Faces in 'Real-Life' Images: Detection, Alignment, and Recognition}, Marseille, France, October 12, (2008).

\bibitem{johnson_mimic-iv_nodate}
A.~Johnson, L.~Bulgarelli, T.~Pollard, B.~Gow, B.~Moody, S.~Horng, L.~A. Celi, and R.~Mark.
\newblock {MIMIC}-{IV} (version 3.0).
\newblock PhysioNet, \url{https://doi.org/10.13026/hxp0-hg59}, last accessed 2024/09/06.

\bibitem{noauthor_nijs_nodate}
J.~Hunt.
\newblock {NIJ}'s recidivism challenge - data.
\newblock NIJ, \url{https://data.ojp.usdoj.gov/stories/s/NIJ-s-Recidivism-Challenge-Data/daxx-hznc/}, last accessed 2024/06/17.

\bibitem{bautista2023ai}
A.~B.~Castillo, A.~Chun, T.~P Vogel, and I.~A Kakadiaris.
\newblock {AI-MET}: A deep learning-based clinical decision support system for distinguishing multisystem inflammatory syndrome in children from endemic typhus.
\newblock {\em medRxiv}, pages 2023--06, (2023).

\end{thebibliography}

\end{document}